\documentclass[CERN,manyauthors]{cernphprep}
\usepackage{float}
\usepackage[comma,square,numbers,sort&compress]{natbib}
\usepackage{hyperref}
\usepackage{lineno}
\usepackage{xspace}
\usepackage{xcolor}
\usepackage{epsfig}
\usepackage{lmodern}
\usepackage{slantsc}
\usepackage{relsize}

\usepackage{booktabs}
\usepackage{siunitx}
\usepackage{multirow}
\usepackage{graphicx}
\usepackage{subcaption}
\def\babar{\mbox{\slshape B\kern-0.1em{\smaller A}\kern-0.1em
    B\kern-0.1em{\smaller A\kern-0.2em R}}}
\def\root{\mbox{\slshape R\kern-0.1em{\smaller OO}\kern-0.1em T~}}

\newcommand*{\epem}{\ensuremath{\mathrm{e^{+}e^{-}}}}

\begin{document}

\begin{titlepage}

\title{
LEP~Data@EDM4hep: mitigating data loss risks by increasing data FAIRness, with a view on FCC-ee
}
\ShortTitle{LEP@EDM4hep}   
\begin{center}
Jacopo Fanini$^{1,2,*}$, Gerardo Ganis$^1$, Marcello Maggi$^3$ \\

{$^1$ CERN, Geneva, Switzerland\\
$^2$ University of Paris-Saclay, Orsay, France\\
$^3$ INFN, Bari, Italy\\
$^*$ Corresponding author: jacopo.fanini@cern.ch} 
\end{center}


\begin{abstract}
The LEP data represents the most precise and highest centre-of-mass energy sample of $\epem$ collision data collected to date. Numerous scientific articles have been published since the conclusion of the experiments, underscoring the ongoing relevance of this dataset and the need to secure its long-term availability according to FAIR data preservation principles. These data could also play a crucial new role in the context of the evaluation of the physics potential of FCC-ee, due to the overlapping centre-of-mass energies, offering a valuable benchmark for detector performance and physics analyses. To fulfill this role, the data should be made available in EDM4hep, the standardized event data format currently developed in the context of the common HEP software ecosystem Key4hep. Migrating to EDM4hep would not only beneficial to future studies but also significantly mitigate the risk of data loss, increase accessibility and interoperability, hence facilitate long-term data preservation. A proof of concept workflow to perform the migration has been developed and successfully applied to ALEPH data.
\vspace{2cm}

\vspace{2cm}

\end{abstract}
\end{titlepage}

\setcounter{page}{2}

\section{Introduction}
\subsection{Data preservation in High-Energy Physics}
\label{sec:dplevels}
The issue of {\it long-term data preservation} of High-Energy Physics (HEP) experiments gained prominence as large-scale experiments, requiring extensive time and resource investments, concluded data collection and faced significant reductions in allocated resources. This challenge primarily affected major experiments of the 1990s, including the LEP, HERA, and Tevatron ones, though some PETRA data long-term preservation efforts had already begun in the context of QCD studies at LEP~\cite{kluthpetra:2003}. Recognizing the importance of data retention, a dedicated study group was established in 2009~\cite{protodphep:2009,protodphep:2012}, followed in 2013 by the formation of the DPHEP collaboration~\cite{dphep:website}. Supported by ten national funding agencies and CERN, and partnered with ICFA, DPHEP continues to serve as a facilitator for knowledge sharing and best practices in data preservation. Its impact is reflected in a series of workshops and reports, the most recent from 2024~\cite{dphepworkshop:2024} and 2023~\cite{dphep:2023}, which highlight the enduring value of preserved data and quantify its scientific contributions through post-data-taking publications\footnote{See for example Figure 1 in~\cite{dphep:2023} and the related discussion.}.

As widely discussed in~\cite{dphep:2023}, in ``data preservation'' the term {\it data} includes not only the digital files resulting from running an experiment, but also the software at all levels (from front-end to visualisation tools), and the documentation produced (publications, technical design reports, internal notes, photographs). Similarly, {\it preservation} does not just mean simply storing the data bits, but also ensuring they are accessible and usable, providing adequate human and computing resources. Overall, the goals of data preservation can be summarized by the FAIR principles~\cite{wilkinson2016fair}: data has to be Findable and Accessible, having e.g. a unique and persistent identifier and being documented with clear metadata, Interoperable, employing formats widely used in the scientific community, and Reusable, with a clear usage license and documentation. Based on these principles, DPHEP has defined four data preservation levels~\cite{dphep:2023}:
\begin{itemize}
    \item Level 1: produce additional information (as plots, tables or metadata) to published results to ease its understanding;
    \item Level 2: preserve some data (e.g. 4-vectors of particles, total energy, etc.) in a simplified form;
    \item Level 3: preserve original data format used for analysis and the software needed to read and process it;
    \item Level 4: preserve data and software to reproduce the full chain, including digitisation and reconstruction, and the ability to generate new simulated events.
\end{itemize}

\subsection{The LEP case and synergies with FCC-ee}
The 2023 DPHEP report~\cite{dphep:2023} also underscores the crucial role of efficient data access in the preparation of future experiments, particularly when they share a common experimental environment. In this context, LEP data has a particular relevance for the electron-positron phase of the proposed Future Circular Collider (FCC) infrastructure~\cite{FCCfeasibility}, which has been recently recommended by the European Strategy Group as the preferred option for the next flagship project at CERN~\cite{CERN-ESU-2025-002}. In its initial phase, FCC will operate with electron-positron collisions at center-of-mass energies that, in part, overlap with those of LEP. Understanding the optimal detector characteristics to fully exploit the statistical potential of FCC-ee’s nominal data samples requires detailed studies. The opportunity to analyse real electron-positron collision data from LEP at similar center-of-mass energies to those foreseen for FCC-ee has therefore garnered significant interest, highlighting the crucial role of LEP data in shaping the future of collider physics.
This data can refine algorithms, reduce uncertainties, and contribute to training the next generation of physicists. Beyond mere preservation, leveraging LEP data in this context represents a pathway for scientific advancement and methodological acceleration.

In addition, the availability of LEP data could be very valuable for LEP scientific output itself, by applying modern, recently developed analyses techniques (including machine learning approaches) to the collected data. Here is a non-exhaustive list where major improvements could be achieved compared to the primary results\footnote{We acknowledge a certain bias toward ALEPH data, as we are most familiar with them. However, there are undoubtedly examples specific to other experiments as well.}:
\begin{itemize}
    \item ALEPH had developed for the $ Z \to hadrons $ total cross-section and the hadron to lepton ratio, a beautiful technique of event rotation, which was statistically as powerful  for the determination of acceptance as a straight Monte Carlo (MC) with two to three order of magnitude more events. This method was not at the time applied for the lepton final states, but this would definitely be of interest for the reduction of uncertainties at FCC-ee; importantly, the application of such a method might lead to interesting results or reduction of errors on ALEPH data themselves.
    \item Similar comments apply to the heavy flavor EW observables, for which the FCC-ee members are constructing new algorithms using LHC machine learning techniques for b- and c-tagging, and possibly for jet charge.
    \item Searches for feebly coupled particles have not been done uniformly for all LEP experiments. This is particularly true for the Heavy Neutral Lepton search in Z decays, for which only DELPHI produced a result. The FCC-ee study group for Long Lived Particles is presently gearing up for developing this search in view of FCC-ee. Clearly, there is a potential for completing and modernizing the LEP output on this topic by producing an ALEPH number, including channels (decays involving taus) which are difficult for the LHC.
    \item Finally, ALEPH was particularly well placed among LEP experiments for tau physics. Tau physics plays a very visible role at FCC-ee, and the availability of ALEPH data will certainly offer a very useful reference point for the development of new algorithms towards FCC-ee. Here again, we observed that ALEPH had not performed a measurement of the tau mass with its own data.
\end{itemize}

\subsection{Status of LEP data}
\label{sec:lepdata}
The latest status of the situation with LEP data is described in~\cite{dphep:2023} and summarized in its essential points in this subsection.
The data collected by the four experiments - ALEPH, DELPHI, L3, OPAL - from 1989 to 2000 were initially stored on CASTOR and later migrated to CTA and CERN’s distributed storage system EOS~\cite{Peters:2134573} for easier access.
With the exception of L3, the LEP experiments have taken specific steps to sustain access to the data, requiring in general specific legacy software, available, as almost all the HEP software, on CernVM-FS~\cite{blomercvmfs:2015}. Containerized solutions techniques for providing older binaries are usually offered; DELPHI and OPAL have also ported the required software to the latest OSes and architectures. L3 decided to migrate some of the ntuple data used for analysis to ROOT files, reducing the need of guaranteeing access to dedicated legacy software. The latter solution is particularly interesting in the context of this work, as it underlines the principle of risk mitigation through use of a common technology (ROOT data format, in this case)~\footnote{It should be noted that the open availability of the L3 data, including the mentioned ntuples, at the time of writing is still to be confirmed.}. 

\subsection{The role of Key4hep and EDM4hep}

\begin{figure}[h]
\begin{center}
\includegraphics[width=12cm]{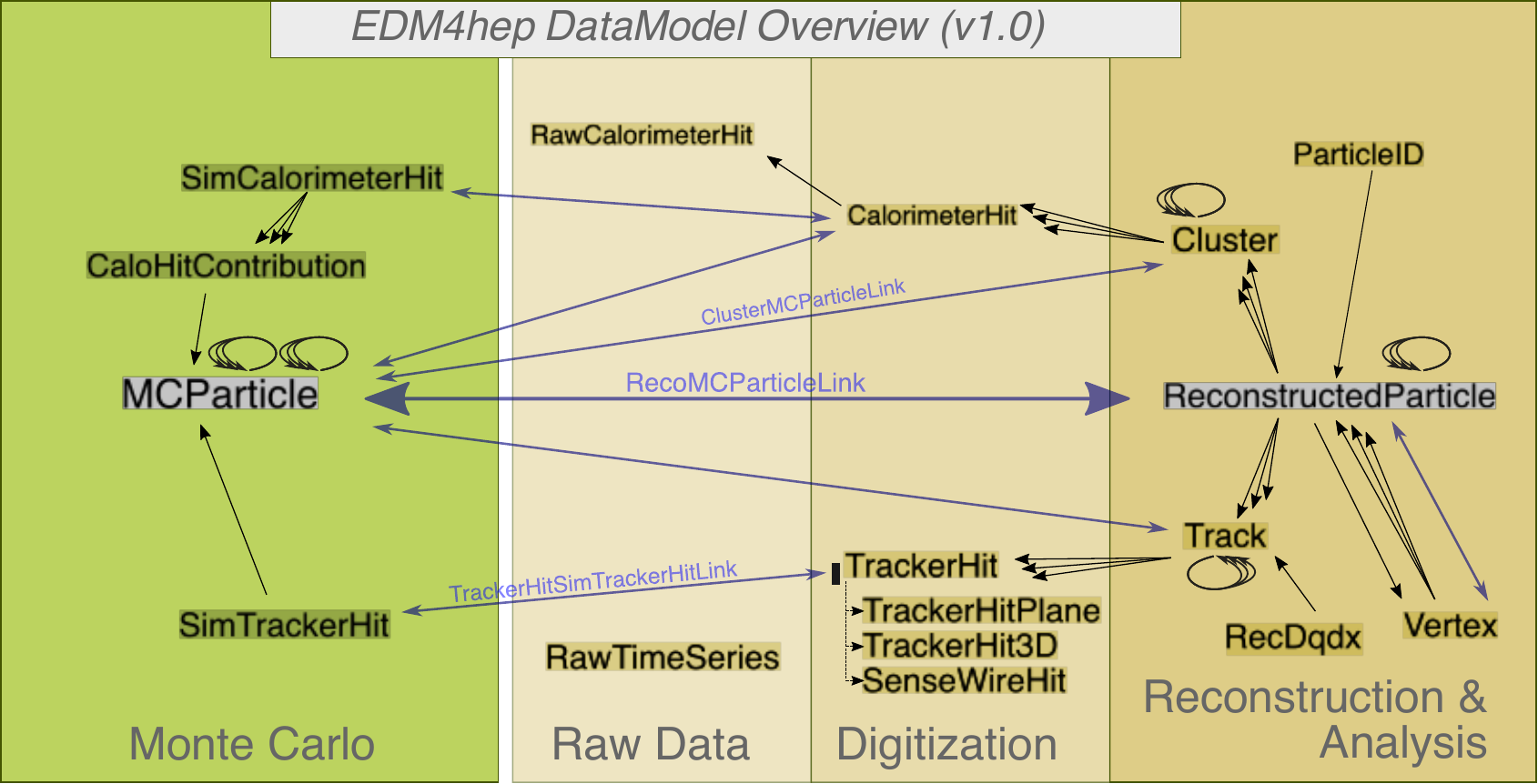}
\end{center}
 \caption{EDM4hep data types schema from simulation to reconstruction. Relations are indicated by the black arrows, while blue arrows indicate external links.}
  \label{figure:edm4hep}
\end{figure}

Key4hep~\cite{Key4hep} is a software ecosystem designed to serve as - and aiming at becoming - a unified framework for high-energy physics experiments. Built upon the experience gained from LHC projects and targeted R\&D initiatives, its goal is to provide a common, standardized access point to established HEP software while offering a framework for developing experiment-specific solutions. The project, developed in the context of future HEP experiments, including FCC, ILC, CLIC, CEPC and Muon Collider, which serve a natural testbed, is now the basis of the FCC software and is being adopted by R\&D activities such the DRDs~\cite{drds:2024}. Key4hep provides a comprehensive framework that covers the entire workflow, from event generation to final data analysis, by including widely used tools, such as Gaudi as event processing framework~\cite{Gaudi}, DD4hep to build the detector's geometries~\cite{dd4hep} and PODIO for its event data model, EDM4hep~\cite{EDM4hep}. The latter is a standardized event data model, aiming to provide a common data structure to organize the processing of the data produced by experiments, both transiently and persistently. EDM4hep design, depicted in Figure~\ref{figure:edm4hep}, consists of several datatypes (for example: ReconstructedParticle, MCParticle, Track), connected together using "internal" relations (e.g. the one associating ReconstructedParticle and Track), or external links (for instance RecoMCParticleLink).

Key4hep and EDM4hep have the potential to be instrumental for both long-term data preservation and the use of LEP data for FCC-ee studies: on the one hand, EDM4hep is already becoming a shared data model across different HEP communities, an essential step for long-term accessibility and interoperability of the migrated data; on the other hand, Key4hep and EDM4hep are the frameworks of choice for FCC-ee studies; thus using them will facilitate the application of already developed tools to analyse the migrated LEP data.

\subsection{Objectives of the work}

The present work describes a pioneering study to strengthen the long-term preservation of LEP data and facilitate its use in FCC studies using as an example the ALEPH archived data. The migration includes also Monte Carlo simulated datasets, along with the relevant metadata. Based on Section~\ref{sec:lepdata}, the aforementioned objectives should be feasible for at least three of the LEP experiments — ALEPH, DELPHI, and OPAL. As mentioned above, it remains unclear whether preservation level 3 data can be recovered for L3. The plan is to first demonstrate the entire process using at least one experiment as a proof of concept. ALEPH data has been chosen as the starting point because it can be seen as the most challenging one, having the software not being ported to the latest OS versions~\footnote{And also because of the familiarity of the authors with the ALEPH environment.}. The general procedure described in this paper should be applicable to other experiments with minor adjustments, and possibly simplified. Discussions have already begun on how to extend this approach to DELPHI and OPAL~\cite{Schwickerath2026_Priv, liko:2026}.

\section{Migrating LEP data to EDM4hep: the ALEPH case}

Data collected by HEP experiments is usually available in experiment-specific formats. In order to migrate it to a different format, it is necessary first to ensure easy access to the data itself. This aspect is not a given, especially when dealing with experiments that stopped collecting data decades ago, as software evolves very rapidly. Moreover, the end of support for the experiment's computing resources, comprising software for data extraction, interpretation, analysis, and visualization, usually follows closely the end of the scientific collaboration. Unless appropriate porting and validation actions are taken, this leads to the freezing of the status of the experiment's software to the last validated computing environment. In addition, accessibility is bound to decay in time in absence of active maintenance. The latter risk is mitigated by making the relevant binaries available on CernVM-FS~\cite{blomercvmfs:2015}, which guarantees persistency and distributed access with a combination of the most consolidated protocols, e.g. POSIX and HTTP. 

\subsection{Computing environment recreation}
For the ALEPH case, the latest version of the code was compiled with Scientific Linux CERN (SLC) 4 (supported until October 2010), with the binaries being also compatible with SLC 6 (CERN Linux production version until 2017, supported until November 2020), the last version validated bit-to-bit by the ALEPH collaboration. Both SLC 6 and ALEPH software are available on CernVM-FS under a dedicated repository {\tt /cvmfs/aleph.cern.ch/}. The source code was managed through various control systems, including HISTORIAN~\cite{historian:1988}, with a final migration to CVS~\cite{cvs:website}, used at CERN until late 2000's. A migration from CVS to Git was performed using {\tt git cvsimport} to have a reference for subroutines and functions.

To overcome the fact that both SLC~4 and SLC~6 are unsupported and non-recommended, an SLC~4/SLC~6 compatible ALEPH computing environment is recreated via virtualization based on the CernVM technology described in Sect.~5.7 of~\cite{dphep:2023}. Images compatible with of all CERN production Linux versions starting with SLC~4, are available on CernVM-FS to serve the CernVM Appliance bootloader technology~\cite{cvmapp:website}.
Container technologies such as {\tt singularity} or {\tt apptainer}~\cite{apptainer} can be used to recreate SLC~4/SLC~6 environment shells from the old OS images available in
CernVM-FS~\footnote{Some modern Linux systems, such as Ubuntu, have dropped 32-bit support in the kernel, preventing the container-based technology to work properly. On these systems the technology can still be used inside a compatible virtual machine created for example with the CernVM Appliance technology and virtualization software such as VirtualBox~\cite{virtualbox}.}.
In the process of recreating these shells, the potentially critical services, such as access to EOS and CernVM-FS, are mounted into the container from the host machine as POSIX mounts points, deferring all critical aspects, such as those related to security, to the host. 
In this way, applications and their dependencies are packaged in an isolated environment that ensures security and consistency across different computing platforms. Examples of scripts to start such shells are available at~\cite{alephsingularitytools}.

Once a compatible environment is recreated, most of the utilities part of the ALEPH offline software system~\cite{Casper:805894} can be used to process the ALEPH data, which at CERN are publicly available under EOS at {\tt /eos/experiment/aleph/}. 

\subsection{Data formats}

ALEPH archived data are stored in different formats, corresponding to different stages of data processing. Each data file contains a run record, describing the run/Monte Carlo conditions during event recording/generation, and, for each event, multiple records, each stored in the form of data banks using the BOS dynamic memory management system~\cite{Blobel:1986, Casper:805894}.

The data formats used by the ALEPH collaboration were:
\begin{itemize}
    \item RAW: information read directly from the detector or generated in case of Monte Carlo data. There is no reconstruction.
    \item POT: result of JULIA\footnote{ALEPH's reconstruction software.}, i.e. tracks, clusters and particles. Every event recorded in the RAW format is also present in POT files.
    \item DST: All POT information, except for noise and background.
    \item Mini-DST: high-level analysis results, sorted in a scaled, integerized and compressed format. This format was used for almost all analyses performed by the ALEPH collaboration.
    \item Nano-DST: most compact format used only for final stages of analyses.
\end{itemize}

Since we target at least DP level 3 (see description in~\ref{sec:dplevels}), the Mini-DST -- the data format used for most ALEPH analyses -- was selected for the migration exercise presented in this paper. The methodology described below can, in principle, be applied to any BOS bank, and thus to any of the data formats listed above.

\subsection{Data conversion methodology and workflows}
\begin{figure}[h]
\begin{center}
\includegraphics[width=13cm]{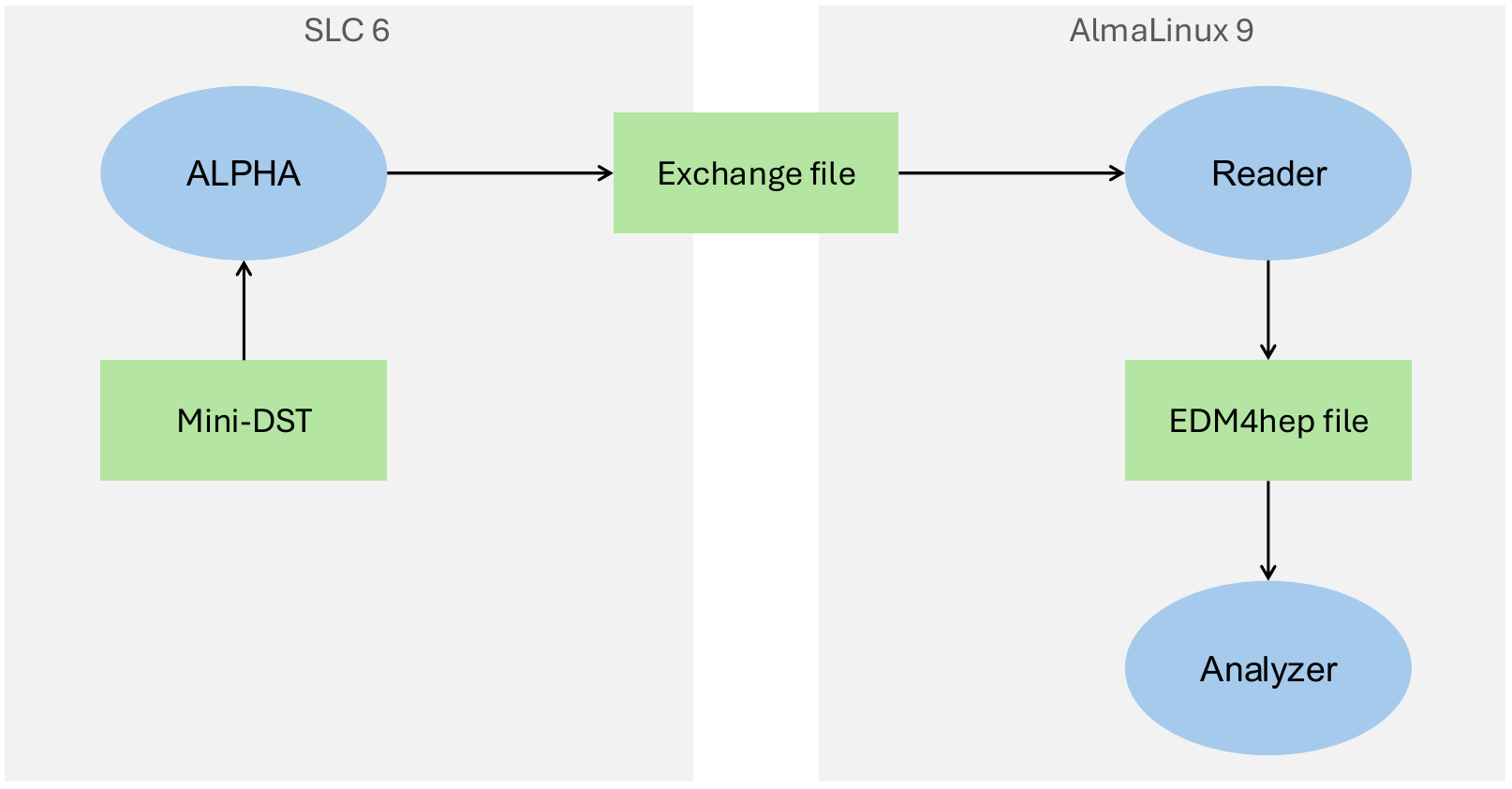}
\end{center}
 \caption{Workflow of the migration process. Green rectangles show the files, blue ellipses the programs developed to perform the data migration. The two computing environments involved in the migration are shown in light grey.}
  \label{figure:workflow}
\end{figure}

The ultimate goal of the workflow is to produce an EDM4hep ROOT file containing the information stored in archived ALEPH files, preserving as much as possible the structure offered by the event data model. Two computing environments are involved:
\begin{itemize}
    \item SLC 6, recreated by emulation, to use the ALEPH software;
    \item AlmaLinux 9, which is the today production computing environment at CERN.
\end{itemize}
The two are interfaced through an exchange file, which is generated in the ALEPH environment and then used on AlmaLinux 9, as is shown in Figure~\ref{figure:workflow}.
Archived ALEPH files, in the Mini-DST format, are available on EOS. Mini-DST files contain most of the data needed for physics analyses, including tracks, vertexes, jets and calorimetric objects, particle identification and the result of ALEPH's particle flow (PF) algorithm~\cite{Haywood:805824}. A Fortran 77 ALPHA\footnote{ALEPH's original physics analyses package.} program is developed to loop event-by-event over the data banks containing the information to be extracted and print the desired data to the exchange file. Some technical issues were encountered and successfully overcome. For instance, the presence of two types of banks (linear and tabular) requires the development of two printing-out functions. Additionally, in Fortran 77 programming language, the output format should be specified in the code, but this detail -- needed because some of the information contained in the banks is stored not as numbers but as letters -- is not available at runtime: the workaround consists in printing all the wanted data as integer values on the exchange file.

The two computing environments are bridged using a simple text file, produced directly by ALPHA routines and easily readable by the program used for the generation of EDM4hep files. In this way, the conversion process is kept as simple and light as possible. On AlmaLinux 9, a C++ program reads the text file and stores in memory all the information that has to be migrated; converting when needed integer values in floats, using the IEEE Standard for floating-point Arithmetic (IEEE-754~\cite{8766229}), or strings. It then fills EDM4hep data structures, creating also relations between associated objects, and once finished it writes the output file. Since the original ALEPH data model and EDM4hep differ, it is not possible to map exactly all the data available in the ALEPH banks to EDM4hep datatypes. Therefore, the conversion is carried out with the subset of data that has a direct correspondence in the new data model. However, EDM4hep allows the introduction of user-defined datatypes to cope with these cases, and this option is under investigation. The output EDM4hep ROOT file stores several objects, including, for instance:
\begin{itemize}
    \item RecoParticles, hosting  the output of ALEPH's PF algorithm (among the others: momentum, energy, mass, charge of each particle);
    \item Tracks, containing the helix fit parameters;
    \item Vertexes, with the position of the primary vertex;
    \item ParticleID, storing the type of the object used to create a certain RecoParticle;
    \item Clusters energy and direction.  
\end{itemize}
For Monte Carlo samples, EDM4hep files host also generator-level information, as the cross section and the list of generated Monte Carlo truth particles.
Relations and links between collections are also in place. Note also that the list above is not exhaustive, and the workflow is designed to be easily adapted to migrate more data, accomodating user requests. The produced files can be analyzed using FCCAnalyses~\cite{fccanalyses2026v0.12.0},a framework developed within the Key4hep ecosystem to deal with the multiple layers of objects introduced by PODIO.

\subsection{Validation of the migrated data}
To ensure consistency between the archived files and the new migrated data, proper validation of the workflow is fundamental.
To achieve this goal, a detailed comparison on a {\it representative} sample of events has been performed. During ALEPH data-taking, events were pre-classified and flagged based on their observable characteristics in a certain number of ``classes'' which, in the following steps, could be easily pre-selected through an event directory mechanism. In this context, the class 15 defining criteria aimed to inclusively select dilepton events, while the class 16 criteria targetted the hadronic events (the detailed list of cuts defining the two classes can be found in Appendix~\ref{app:classes}). The sample obtained from the union of classes 15 and 16 therefore includes the events used, for example, in electroweak studies and can, for the purposes of this work, be considered a {\it representative} sample for validation.

The workflow is validated reproducing these cuts in the two computing environments, using on one side ALPHA routines and archived files and on the other FCCAnalyses and the migrated dataset. Then, a comparison of the events selected in the two cases is conducted. A subset of around 350k events from 1994 data is used for the validation, resulting in less than 0.002 \% of mis-selected events, i.e. events selected by only one of the two algorithms in the two different computing environments.

A detailed investigation reveals that the reason why some events are not selected by one of the two algorithms is ultimately related to the presence of correction algorithms applied when ALPHA is used, which cannot be easily replicated when applying cuts on the migrated data. These corrections, of the order of a few parts per thousand, slightly modify some of the measured values, leading to erroneous non-selection of few events. An example is shown in Table~\ref{tab:mom-comparison}: the magnitude of the momentum of the second particle changes from $3.005\,\mathrm{GeV}/c$ in ALPHA to $2.994\,\mathrm{GeV}/c$ in FCCAnalyses; since the selection algorithm accepts only events with at least one particle having total momentum greater than  $3\,\mathrm{GeV}/c$, the ALPHA routine selects the event, while the FCCAnalyses function discards it. Additionally, the two routines may not be exactly the same, as some operations, e.g. the thrust computation, are performed using pre-defined routines that might not perfectly overlap.

\begin{table}[htbp]
\centering
\caption{Momentum components computed with ALPHA and FCCAnalyses applying class 15 selection algorithms to the same event. Each line corresponds to a different particle. The total momentum of the second particle is shown in bold to underline the difference causing the mis-selection error.}
\label{tab:mom-comparison}
\begin{tabular}{l c c c c}
\toprule
Sample & $p_x$ [GeV/c] & $p_y$ [GeV/c] & $p_z$ [GeV/c] & $p_T$ [GeV/c] \\
\midrule
\multirow{2}{*}{ALPHA}
 & -0.170 & -0.127 & 0.233 & 0.315 \\
 & 0.662  & 1.694  & -2.392 & \textbf{3.005} \\
\specialrule{1.5pt}{0pt}{0pt}
\multirow{2}{*}{FCCAnalyses}
 & -0.169 & -0.127 & 0.232 & 0.314 \\
 & 0.659  & 1.688  & -2.384 & \textbf{2.994} \\
\bottomrule
\end{tabular}
\end{table}

 Validation results are summarized in Table~\ref{tab:valid}. As the number of events affected by this issue would correspond to a systematic effect orders of magnitude smaller than the statistical power of the samples, the validation is considered successful at this demonstrative stage. The absence of software corrections in the migrated data is also seen as an advantage, as it preserves a closer representation of the original dataset. Nevertheless, the migration workflow remains flexible and could be adapted at a later stage to extract different variants of the archived data if required for further analysis.

\begin{table}[htbp]
\centering
\caption{Results of the validation. ALPHA-only column shows the number of events selected as good events from the ALPHA routine using the archived dataset, but not from the same selection algorithm running on the migrated dataset. Conversely, FCC-only column collects the number of events passing the cuts only in the migrated dataset.}
\label{tab:valid}
\begin{tabular}{llcc}
\toprule
Events & Selection & ALPHA-only & FCC-only \\
\midrule
\multirow{2}{*}{367501}
 & class 15 & 5 (0.0014\%) & 0 (0\%) \\
 & class 16 & 4 (0.0011\%) & 1 (0.0003\%) \\
\bottomrule
\end{tabular}
\end{table}

\begin{figure}[h]
\begin{center}
\includegraphics[width=13cm]{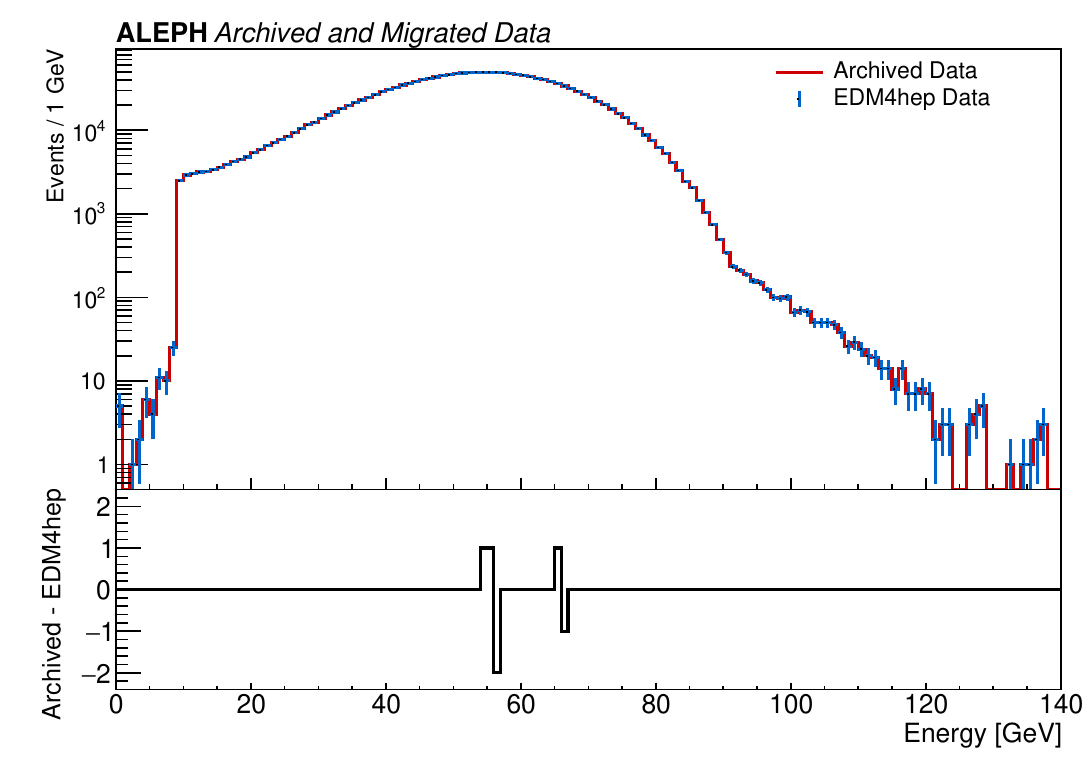}
\end{center}
 \caption{Charged particles' total energy distribution in 1994 class 16 selected events. Data migrated to EDM4hep are shown as blue markers, while archived data are shown as a red line. The lower panel shows the difference in the number of events in each energy bin between archived and EDM4hep data.}
 \label{figure:originalvsmigrated}
\end{figure}

The next step in validating the workflow is the comparison of key observables obtained from the migrated EDM4hep files with the same ones, extracted from the archived files. An example of this comparison is shown in Figure~\ref{figure:originalvsmigrated} for the total energy of class 16 charged particles. As can be seen in the Figure, the agreement between the two distributions is almost total and compatible with results shown in Table~\ref{tab:valid}, additionally consolidating the confidence in the data migration chain.

\begin{figure}[h]
\begin{center}
\includegraphics[width=13cm]{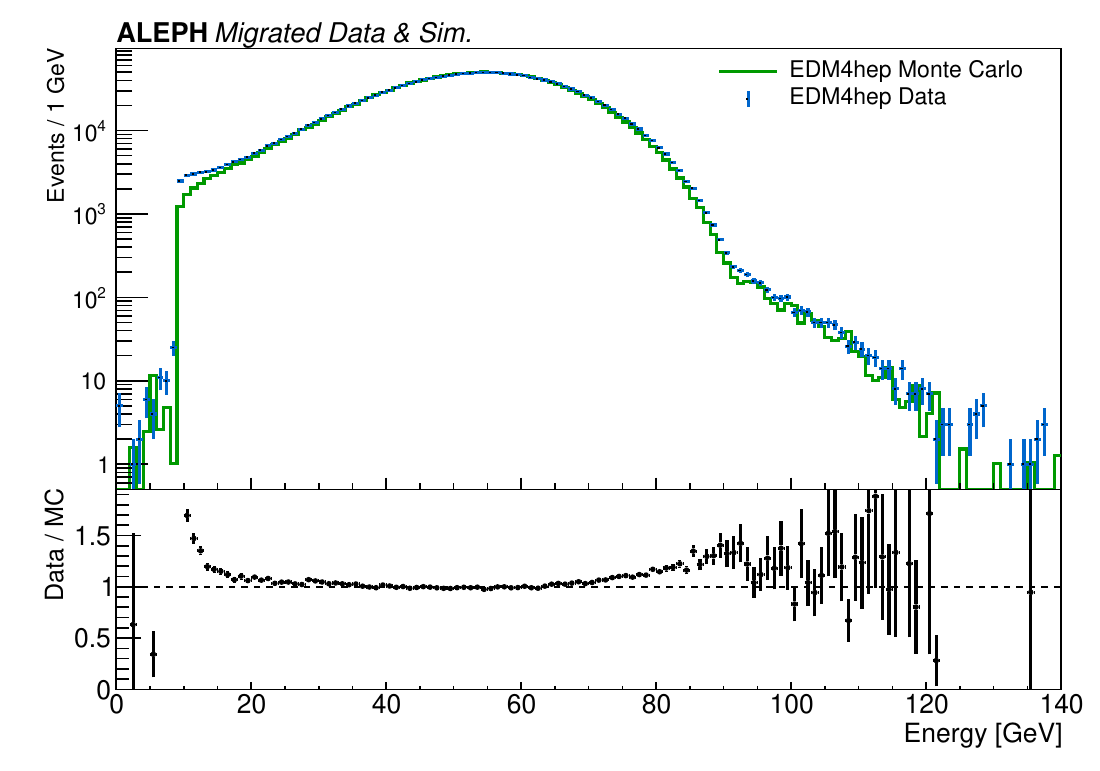}
\end{center}
 \caption{Charged particles' total energy distribution in 1994 class 16 selected data and $q\bar{q}$ simulated data EDM4hep samples. Data are shown as blue markers, while the Monte Carlo simulation is shown as a green line. The lower panel shows the ratio between data and Monte Carlo for each energy bin.}
  \label{figure:migrated}
\end{figure}

Beyond the comparison between archived and migrated datasets, a comparison between real and simulated data after migration to EDM4hep is shown in Figure~\ref{figure:migrated}. The agreement beetwen the two distributions is at the level expected~\footnote{See for example Figure~1 in~\cite{alephew:1992}.}, in particular considering the fact that the Monte Carlo sample includes only the dominant $q\bar{q}$ processes.

\newpage
\section{Status of the ALEPH data migration to EDM4hep}
\subsection{Data availabilty and documentation}
The produced ROOT files are stored on EOS and are publicly available under

{\centering{\tt /eos/experiment/aleph/EDM4HEP/}.\par}

The migration of the dataset has been performed for 1994 dataset, the single dataset with most of the data collected at the Z peak.  Monte Carlo events are also available for the hadronic final states, along with a database with the relevant metadata. The size and number of files for archived and migrated datasets are summarized in Table~\ref{tab:datasets}. Additional events from different data-taking years can be migrated with a minimum overhead.

\begin{table}[htbp]
\centering
\caption{Summary of archived and migrated 1994 datasets. For the data samples, an EDM4hep file is created for each "fill" of the LEP machine (which usually has several runs), in this way events in the same file should have the same data-taking conditions.}
\label{tab:datasets}
\begin{tabular}{l l c c}
\toprule
Sample & Category & Number of files & Total size (GB) \\
\midrule
\multirow{2}{*}{Data}
 & Archived        & 63  & 19 \\
 & Migrated & 189  & 18 \\
\specialrule{1.5pt}{0pt}{0pt} 
\multirow{2}{*}{MC ($ q\bar{q}$ events)}
 & Archived        & 70 & 19 \\
 & Migrated & 70  & 11 \\
\bottomrule
\end{tabular}
\end{table}

The available documentation is hosted on the website

\begin{figure}[h]
\begin{center}
\fbox{\includegraphics[width=12cm]{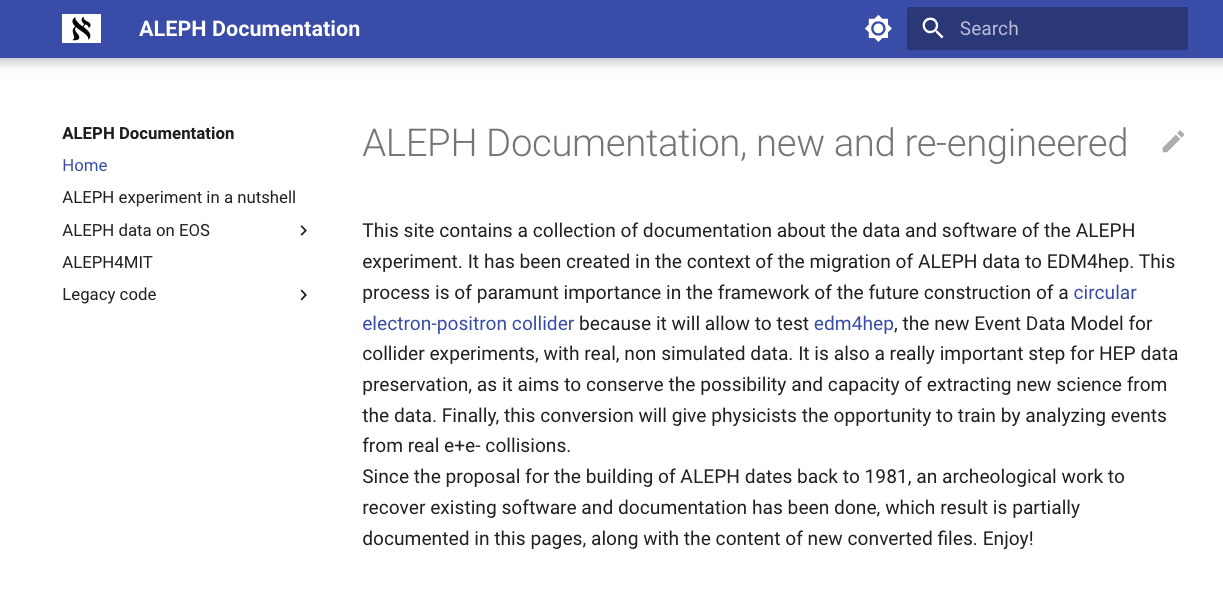}}
\end{center}
 \caption{Screenshot of the website home page documenting, among other things, how to recreate ALEPH's computing environment and the content of migrated files.}
  \label{figure:website}
\end{figure}

{\centering{\tt https://aleph-new.docs.cern.ch/},\par}

accessible from within the CERN domain (after CERN login). Figure~\ref{figure:website} shows the homepage of the website. In particular, the documentation provides a description of the content of the new migrated files. 

\subsection{First look at the migrated data}

ALEPH migrated data are already being re-analyzed by some users, highlighting the interest of the community in having them available in an easily accessible format. The most advanced results refer to the application of modern jet flavor tagging machine learning algorithms, such as ParticleNet~\cite{Qu_2020} or ParticleTransformer~\cite{qu2024particletransformerjettagging}, to sharpen the separation between the various hadronic decay channels of the Z boson. Compared to legacy ALEPH results, modern taggers are achieving roughly an order of magnitude improvement in both light- and c-jets background rejection at the same efficiency level~\cite{defranchis:2026}. These very promising results open the door to more accurate measurements of electroweak precision observables using LEP data.

\section{Summary and outlook}
This work describes a first-of-a-kind approach to data preservation, trying to migrate data stored in a no-longer easily usable data format to a persisted and supported ROOT-based event data model, while keeping open the possibility to use the experiment's original software.
LEP data represent a unique and invaluable sample of $\epem$ collision events, and its migration to a common format accessible via experiment-agnostic tools comes at a minimal cost while offering a potentially high return-on-investment. This effort will significantly enhance data availability for FCC preparation studies and, importantly, contribute to training the next generation of physicists who will fully exploit the potential of FCC-ee.

Future developments are foreseen and already planned, including:
\begin{itemize}
    \item Enlarge the range of extracted data, thus exploiting as much as possible all the information stored in the original archived files. This requires a dialog (already ongoing) with the users to understand which features are needed for FCC-ee studies. Feedback from the community is also essential to further validate the workflow, as errors and bugs are inevitable. Once a satisfactory level of information is reached in the EDM4hep ROOT file, the migration will be extended to all years of data-taking.
    \item Establish a methodology for recreating Monte Carlo datasets using modern or post-LEP versions of event generators. This presents both a challenge and an opportunity: leveraging the existing infrastructure to integrate modern Monte Carlo generators and physics models with the legacy LEP simulation software stacks, primarily based on Geant3. By propagating four-vectors through these well-established detector descriptions and tuning the models directly on LEP data, it becomes possible to significantly enhance the accuracy and predictive power of FCC-ee era simulations well before FCC-ee begins data collection. This approach offers a unique opportunity to optimize detector response modeling and refine analysis strategies in advance, ensuring that FCC-ee can fully exploit its scientific potential from the outset.
\end{itemize}

In the longer term, extending this approach to other LEP experiments could lay the foundation for a more sustainable strategy in preserving HEP data—particularly if the critical LEP expertise still available is captured during the process. There is no doubt that making these datasets functional to the FCC-ee physics potential studies would offer an effective pathway to achieving this goal.

\newpage
\bibliographystyle{utphys}   
\bibliography{bibliography}

\newpage
\appendix
\section{Class 15 and class 16 selection algorithms}
\label{app:classes}
The selection algorithm for dilepton candidate events (class 15) proceeds as follows:
\begin{enumerate}
    \item Only tracks with more than 3 hits in the Time Projection Chamber (TPC), $|Z_0| < 10\,\mathrm{cm}$ and 
    $|\mathbf{p}| > 0.1\,\mathrm{GeV}/c$ are used.  
    Their numbers are counted separately for $|D_0| < 5\,\mathrm{cm}$ and 
    $|D_0| < 2\,\mathrm{cm}$.

    \item 
    \begin{enumerate}
        \item If there are exactly 2 tracks with $|D_0| < 5\,\mathrm{cm}$, they are declared as good tracks and the selection continues.
        \item If there are between 2 and 8 tracks with $|D_0| < 2\,\mathrm{cm}$, they are declared as good tracks and the selection continues.
    \end{enumerate}

    \item The thrust axis is calculated, and each of the two hemispheres defined by the axis is required to contain at least one good track.

    \item At least one track with $|D_0| < 2\,\mathrm{cm}$ must have 
    $|\mathbf{p}| > 3\,\mathrm{GeV}/c$.

    \item If there are more than 4 good tracks, each of them is required to have an opening angle with respect to the axis of the corresponding jet satisfying 
    $\cos(\eta) > 0.85$.
\end{enumerate}

An event is flagged as hadronic (class 16) if it satisfies the following conditions:
\begin{enumerate}
    \item There are at least 5 tracks, each with more than 3 hits in the TPC, satisfying the following cuts:
    \begin{itemize}
        \item $|D_0| < 2\,\mathrm{cm}$,
        \item $|Z_0| < 10\,\mathrm{cm}$,
        \item $\cos(\theta) < 0.95$.
    \end{itemize}

    \item The total energy of the selected tracks must be greater than $10\%$ of the center-of-mass energy.
\end{enumerate}

\end{document}